\documentclass[aps,prb,twocolumn]{revtex4}

\usepackage{graphics}
\usepackage{epsfig}

\usepackage{color}

\def\bra#1{\left\langle#1\right|}
\def\ket#1{\left|#1\right\rangle}

\begin{document}

%Include all the eps files in Figures directory
\graphicspath{{Figures/}}

\title{Axion topological field theory of topological superconductors}

\author{Xiao-Liang Qi$^1$}
\author{Edward Witten$^2$}
\author{Shou-Cheng Zhang$^1$}
\affiliation{
$^1$Department of Physics, Stanford University, Stanford, CA 94305, USA\\
$^2$School of Natural Sciences, Institute for Advanced Study, Princeton NJ 08540
}

\date{\today}

\begin{abstract}
Topological superconductors are gapped superconductors with gapless and topologically robust quasiparticles propagating on the boundary. In this paper, we present a topological field theory description of  three-dimensional time-reversal invariant topological superconductors. In our theory the topological superconductor is characterized by a topological coupling between the electromagnetic field and the superconducting phase fluctuation, which has the same form as the coupling of ``axions" with an Abelian gauge field. As a physical consequence of our theory, we predict the level crossing induced by the crossing of special ``chiral" vortex lines, which can be realized by considering $s$-wave superconductors in proximity with the topological superconductor. Our theory can also be generalized to the coupling with a gravitational field.
\end{abstract}

%\pacs{05.10.Cc,75.10.Jm,71.10.-w}

\maketitle

%Outline:
%
%1. TI and topological field theory description. 3d Topological superconductor. Several effective theory descriptions. Comment on the problem of gravity theory. Absence of field theory description.
%2. He3B + SOC (lattice model). Comparison of topological trivial and nontrivial SC. Generalization to generic winding number N.
%3. 4+1d regularization. Axion field theory. $N=1$ case and general case.
%4. Consequence of the axion field theory. (1) $s$-wave proximity effect. (2) vortex line linking.

\section{Introduction}

Since the discovery of the quantum Hall effect in the 1980's\cite{klitzing1980,tsui1982}, the study of topological states of matter (TSM) has
been an active field of research in modern condensed matter physics. TSM are those states of matter which are distinguished from each other by some topological properties rather than more conventional properties such as symmetries preserved or broken by the order parameter. Recently, wide classes of TSM's known as topological insulators (TI) and topological superconductors (TSC) have been proposed, and TI's have been experimentally realized in various materials\cite{qi2011rmp,hasan2010,moore2010}. TI's are insulators with a bulk gap and gapless edge/surface states propagating on the boundary which are robust against perturbations within the given symmetry class. Similarly, TSC's are fully gapped superconductors with a bulk gap for the quasi-particle excitations, and topologically protected gapless quasi-particle states propagating on the boundary. The first example of topological superconductor is the $(p+ip)$-wave superconductor of spinless fermions in two-dimensions\cite{read2000}, which has a chiral Majorana fermion state propagating on the edge. (The Majorana fermion of the edge state refers to the fact that the quasiparticle is its own anti-particle.) More recently, a new class of TSC was proposed in three-dimensional (3d) time-reversal invariant (TRI) superconductors\cite{schnyder2008,roy2008,qi2009b}, which has spin-triplet pairing in the bulk, and two-dimensional massless Majorana surface states with linear ``relativistic'' dispersion. The $^3$He B phase was proposed as an example of the 3d TRI TSC (more rigorously, a topological superfluid)\cite{chung2009}. Solid state candidates of 3d TSC have also been proposed\cite{yan2010a,fu2010b}. A recent tunneling experiment\cite{sasaki2011} supports the theoretical proposal of Cu doped ${\rm Bi_2Se_3}$ as a TSC\cite{fu2010b}, although more experiments are needed to reach a conclusion.

Without interactions, TI and TSC can be described by topological band theories\cite{qi2011rmp,hasan2010}, which characterize a class of TI or TSC by a topological invariant of the band Hamiltonian. The classification of non-interacting TI and TSC is understood systematically in generic dimensions\cite{qi2008b,schnyder2008,kitaev2009}. However, in reality all electron systems are interacting, so that the physically interesting TSM are those which are robust even with interaction. In general it is difficult to directly study the topological classification of gapped interacting Hamiltonians or their ground states, except in some special cases such as in one dimension\cite{fidkowski2010,fidkowski2011b,turner2011,chen2011}, and in some special classes of models in higher dimensions\cite{qi2012,ryu2012,yao2012,chen2011b,gu2012}. A general approach to characterize interacting TSM's is by writing down the possible topological response theories, which describe some observable physical properties of the system, and contain  a``topological order parameter'' that is required to be quantized by general principles. For example, the quantum Hall effect can be characterized by Chern-Simons theories\cite{zhang1989,wen1990b,zhang1992}. The topological order parameter is the coefficient of the Chern-Simons theory which has the physical meaning of Hall conductivity and is quantized due to charge quantization. In the case of 3d TRI TI, the topological response theory is described by a term in the electromagnetic response\cite{qi2008b} $\frac{\theta}{32\pi^2}\int d^4xF_{\mu\nu}F_{\sigma\tau}\epsilon^{\mu\nu\sigma\tau}$, in which $\theta$ is quantized to be $0$ or $\pi$ mod $2\pi$ if the system is time-reversal invariant. This is an example of a topological response theory in which the quantization of the topological order parameter depends on discrete symmetry (time-reversal symmetry in this case). The physical consequence of this topological response theory is the topological magneto-electric effect\cite{qi2008b,qi2009}. Such a topological term has the same form as the coupling of axions with gauge field proposed in high-energy physics\cite{peccei1977,wilczek1987}. However, in a TI, $\theta$ is a constant determined by the bulk topology rather than a dynamical field.

It is therefore natural to ask what is the topological response theory description of TSC. In a superconductor the electromagnetic field is screened, and the edge/surface states of TSC are quasi-particles without a well-defined charge. Therefore the electromagnetic response has been considered as improper for defining the topological response theory. For this reason, the topological BF theory has been proposed to describe the chiral topological superconductor\cite{hansson2011}. Alternatively, the coupling to gravitational field and quantized thermal response have been proposed\cite{read2000,wang2011,ryu2012b}. Similar to the axionic coupling to the electromagnetic field in 3d TI, a term proportional to the Pontryagin invariant of the Riemann curvature is proposed to characterize the gravitational response of 3d TSC, with coefficient $\theta=0$ or $ \pi$ mod $2\pi$\cite{ryu2012b,wang2011}. However, Ref. [\onlinecite{wang2011}] pointed out that the coefficient $\theta$ of the Pontryagin invariant is a $U(1)$ phase, so that under the condition of time-reversal symmetry there are only two physically different values of $\theta$, leading to a $Z_2$ classification of the TSC rather than the integer classification in the non-interacting case. Ref. [\onlinecite{wang2011}] proposed an alternative approach of defining a topological response theory by considering proximity effects of TSC to trivial $s$-wave superconductors. On the surface of a TRI TSC, two thin films of trivial $s$-wave superconductors are coupled to the TSC by the proximity effect and also coupled with each other through Josephson coupling. When the two films of the $s$-wave superconductors have a $\pi$ phase difference, the number of chiral modes propagating along the Josephson junction provides a definition of the integer valued topological invariant of the bulk TSC. This proposal provides a physically operational definition of the topological invariant, which answers the question of whether the integer valued topological invariant of 3d TRI TSC is robust upon electron interaction. However, the $s$-wave proximity effect is proposed as an experimental setting and is not formulated in terms of a topological field theory description.

In this paper, we develop the topological field theory description of 3d TSC by considering the TSC coupled to both the electromagnetic field and the superconducting phase fluctuations. We demonstrate that the electromagnetic response of the TSC contains a topological term
\begin{eqnarray}
\mathcal{L}_{\rm topo}=\sum_i\frac{C_{1i}\theta_i}{64\pi^2}\epsilon^{\mu\nu\sigma\tau}F_{\mu\nu}F_{\sigma\tau}
\label{axion}
\end{eqnarray}
with $i$ labeling all Fermi surfaces and $\theta_i$ the superconducting phase of each Fermi surface. $C_{1i}$ denotes the Chern number describing the spin helicity on each Fermi surface. We dub this theory the ``axion field theory" of TSC since it has the form of the coupling between axion and Abelian gauge theory\cite{peccei1977,wilczek1987}. The details of the derivation and physical consequences will be presented in the later part of this paper. The form of this topological field theory appears similar to that of TI, but there are two essential differences: i) The phases $\theta_i$ are dynamical variables describing the superconducting phases, which are thus coupled to the gauge field $A_\mu$ by gauge coupling, while in TI the phase is a constant topological order parameter; ii) In a superconductor there is a Higgs term for the gauge field $A_\mu$, so that the gauge field is massive, while in TI the gauge field remains gapless in the long wavelength limit. Due to these two important differences, the physical consequences of this topological term are very different from those of the TI case. Due to the Higgs term, the electromagnetic field vanishes in the superconductor except in vortex cores (if the superconductor is type II). For a superconductor with multiple Fermi surfaces, we can consider vortices of only one Fermi surface, which we call chiral vortices since each Fermi surface with a nontrivial Chern number is equivalent to a Weyl fermion with chirality. We show that the topological term (\ref{axion}) describes an anomaly when chiral vortices cross each other, which can be interpreted as the change of ground state fermion number parity caused by level crossing between quasi-particle energy levels. This is an analog of the $Z_2$ Witten anomaly in a system of odd flavors of Weyl fermions coupled to an $SU(2)$ gauge field\cite{witten1982}. The $s$-wave proximity effect proposed in Ref. [\onlinecite{wang2011}] is shown to be a physical realization of the chiral vortices. Therefore our topological field theory correctly describe the proposed $s$-wave proximity effect\cite{wang2011}.

%TI should be described by topological field theory in order to be generally defined. 3dTI described by the $\theta F\wedge F$ theory. Corresponds to the axion in high energy physics. However the axion is not a dynamical field and does not carry gauge charge.
%
%The axion theory with axion carrying gauge charge emerges as effective theory of TSC. The axion has the physical meaning of SC phase in a TSC. The theory explains the puzzle of Z2 vs Z. Explains the previously proposed proximity effect. Physical consequences.

The rest of the paper is organized as follows. The single particle description of 3d TRI TSC is reviewed in Sec. \ref{sec:reviewTSC}, which provides a starting point of our discussion. In Sec. \ref{sec:axion} we present the derivation of the axion field theory by introducing a mapping from the 3d TSC to a four-dimensional TSC. In Sec. \ref{sec:consequence} we discuss the physical consequences of the axion field theory, including the chiral vortex linking effect and the $s$-wave proximity effect. Finally, Sec. \ref{sec:summary} is devoted to the summary and further discussions.

\section{The $(3+1)$-d topological superconductors}\label{sec:reviewTSC}

\subsection{The topological invariant of $(3+1)$-d topological superconductors}

As a starting point of our theory, we review the single particle description of the $(3+1)$-d topological superconductors. At the level of Bogoliubov-de Gennes (BdG) theory, the topological superconductors are described by the generic BdG Hamiltonian\cite{schnyder2008}
\begin{eqnarray}
H=\frac12\sum_{\bf k}\left(c_{\bf k}^\dagger,c_{-\bf k}\right)\left(\begin{array}{cc}h_{\bf k}&\Delta_{\bf k}\\\Delta_{\bf k}^\dagger &-h_{-\bf k}^T\end{array}\right)\left(\begin{array}{c}c_{\bf k}\\c_{-\bf k}^\dagger\end{array}\right)\label{Hbdg}
\end{eqnarray}
with $c_{\bf k}$ a $N$-component vector for a system with $N$ bands, and $h_{\bf k},\Delta_{\bf k}$ $N\times N$ matrices. Time reversal symmetry acts as
\begin{eqnarray}
T\left(c_{\bf k}\right)=\mathcal{T}c_{-\bf k},~T\left(c_{-\bf k}^\dagger\right)=\mathcal{T}^*c_{\bf k}^\dagger\label{Tdef}
\end{eqnarray}
with $\mathcal{T}$ the time-reversal matrix satisfying $\mathcal{T}^T=\mathcal{T}^*=-\mathcal{T},~\mathcal{T}^2=-1$. The time-reversal invariance requirement to the Hamiltonian is
\begin{eqnarray}
\mathcal{T}^{\dagger}h^*_{\bf k}\mathcal{T}=h_{-\bf k},~\mathcal{T}^{\dagger}\Delta_{\bf k}^*\mathcal{T}=-\Delta_{-\bf k}\label{Tcondition}
\end{eqnarray}
By defining ${ Q}_{\bf k}=h_{\bf k}+i\Delta_{\bf k}$, one can prove that $Q_{\bf k}$ is non-singular for gapped superconductors. The integer-valued topological invariant describing three-dimensional topological superconductor is the winding number of the map from the Brillouin zone torus $T^3$ to the special linear group $SL(N,{\rm C})$ defined by $Q_{\bf k}$.

For the purpose of this paper, it is helpful to start from an alternative formula of the topological invariant proposed in Ref. [\onlinecite{qi2010b}] in the weak pairing limit. When the pairing is weak compared to the kinetic energy $h_{\bf k}$, only the pairing near Fermi surfaces of the kinetic energy $h_{\bf k}$ is important for the topological properties of the system. For each Fermi surface $\Sigma_n$, the momentum
dependence of the states $\ket{n{\bf k}}$  on the fermi surface  defines a Berry connection $a_i=-i\bra{n{\bf k}}\partial_{k_i}\ket{n{\bf k}}$, which has a Chern number defined by
\begin{eqnarray}
C_{1n}=\frac1{2\pi}\int_{\Sigma_n}d\Omega^{ij}\left(\partial_ia_j-\partial_ja_i\right)
\end{eqnarray}
If one defines the pairing order parameter near the Fermi surface by
\begin{eqnarray}
\Delta_{n{\bf k}}=T\left(\bra{n{\bf k}}\right)\Delta_{\bf k}\ket{n{\bf k}}\label{pairingsign}
\end{eqnarray}
it can be proved that $\Delta_{n{\bf k}}$ is real due to time-reversal symmetry. The choice of $T(\bra{n{\bf k}})$ as the wavefunction at momentum $-{\bf k}$ avoids the ambiguity in the gauge choice of the wavefunction, and makes the statement of a ``real" pairing $\Delta_{n{\bf k}}$ meaningful, which is only possible because of the time-reversal symmetry. The gap of the BdG Hamiltonian near the Fermi surface is $\left|\Delta_{n{\bf k}}\right|$, so that in a fully gapped superconductor,
$\Delta_{n{\bf k}}$ cannot change sign as a function of $\bf k$ on each Fermi surface. Consequently, in a fully gapped TRI superconductor, the sign of the pairing is definite in each Fermi surface. It was proved in Ref. [\onlinecite{qi2010b}] that the topological invariant is given by the following sum of Fermi surface Chern numbers weighted by the sign of the pairing:
\begin{eqnarray}
N=\frac12\sum_nC_{1n}{\rm sgn}\left(\Delta_{n{\bf k}}\right)\label{TopoInv}
\end{eqnarray}

As an example, we can consider the following two-band model defined by
\begin{eqnarray}
h_{\bf k}&=&\frac{{\bf k}^2}{2m}-\mu+\alpha \sigma\cdot {\bf k}\nonumber\\
\Delta_{\bf k}&=&i\Delta_0\sigma_y\sigma\cdot {\bf k}\label{Hminimal}
\end{eqnarray}
in Eq. (\ref{Hbdg}). The time-reversal transformation is defined by $\mathcal{T}=i\sigma_y$. Up to a basis transformation, the Hamiltonian for $\alpha=0$ describes the He$^3$ BW phase\cite{vollhardt1990,qi2009b,schnyder2008,roy2008}. We add a term $\alpha$ to lift the degeneracy between the two Fermi surfaces, which is convenient for defining the pairing order parameter on the Fermi surface. We would like to note that this splitting is just for convenience and is not physically required.

The two Fermi surfaces are spherical with the Fermi momentum
\begin{eqnarray}
k_{F\pm}=\mp m\alpha+\sqrt{m^2\alpha^2+2m\mu}.
\end{eqnarray}
The states around the two Fermi surfaces are defined by the two eigenstates of $\sigma\cdot{\bf k}$, denoted by $\sigma\cdot{\bf k}\ket{{\bf k}\pm}=\pm|{\bf k}|\ket{{\bf k}\pm}$. It can be verified that the two Fermi surfaces with the opposite spin helicities have opposite Chern numbers and also opposite signs of the pairing order parameter, so that the topological invariant is given by
\begin{eqnarray}
N=\frac12\left[1-(-1)\right]{\rm sgn}\Delta_0={\rm sgn}\Delta_0
\end{eqnarray}

\subsection{A global ambiguity in the definition of the topological invariant}

%XLnote:
%({\bf I added a subsection here, to make the comment about the sign ambiguity.})

Before proceeding to the next section, we would like to make a comment about the definition of the topological invariant (\ref{TopoInv}). The pairing order parameter $\Delta_{n{\bf k}}$ is defined in Eq. (\ref{pairingsign}), which depends on the definition of the time-reversal transformation in Eq. (\ref{Tdef}). If we substitute $\mathcal{T}$ by $-\mathcal{T}$ in Eq. (\ref{Tdef}), all properties of $\mathcal{T}$ are still preserved, and the time-reversal condition for Hamiltonians given by Eq. (\ref{Tcondition}) also remains  valid. Therefore a time-reversal invariant Hamiltonian remains so in the new definition of time-reversal. However, $\Delta_{n{\bf k}}$ defined in Eq. (\ref{pairingsign}) changes sign for all bands. Therefore the topological invariant $N$ is transformed to $-N$. Since changing the sign of $\mathcal{T}$ is equivalent to following the time-reversal transformation by a global phase rotation $c_{\bf k}\rightarrow -c_{\bf k}$, there is no physical way to distinguish the two definitions of time-reversal symmetry. Consequently, topological superconductors with topological invariant $N$ and $-N$ are physically equivalent. However, such an ambiguity only occurs for a global sign of the topological invariant and does not affect any physical consequence of the topological invariant such as surface states at the interface between two superconductors with different topological invariant. It also does not affect the topological field theory and topological defects discussed in the rest of this paper.

\section{$(4+1)$-dimensional regularization and the derivation of the axion field theory}\label{sec:axion}

With the background prepared in the last section, now we show how the theory reviewed above is related to another topological state of matter, the $(4+1)$-dimensional time-reversal invariant topological insulators\cite{zhang2001,qi2008b}. Such a relation will be essential in the axion field theory that we will derive.

\subsection{$(4+1)$-dimensional model of the $(3+1)$-d TSC}

From the Fermi surface formula, it can be seen that Fermi surfaces with nontrivial Chern number are essential for topological superconductivity. On the other hand, a Fermi surface with non-trivial Chern number $C_1=1$ is topologically equivalent to a Weyl fermion (with the fermi level generally away from the Dirac point). For example, in the minimal model (\ref{Hminimal}), the low energy effective theory near each Fermi surface is a Weyl fermion since it has a fixed spin helicity $\sigma\cdot {\bf k}/|k|$. The Fermi surface at $k_{F+}$ has spin helicity $+1$ with a uniform pairing $\Delta_0k_{F+}$. Thus the low energy states around the spin helicity $+1$ Fermi surface in the weak pairing limit is equivalent to the following Weyl fermion theory:
\begin{eqnarray}
H_R&=&\sum_{\bf k}\psi_{R{\bf k}}^\dagger\left[v_{F+}\sigma\cdot{\bf k}-\mu_+\right]\psi_{R{\bf k}}\nonumber\\
& &+\frac12\left(\sum_{\bf k}\Delta_0k_{F+}\psi_{R-{\bf k}}^\dagger i\sigma_y\psi_{R{\bf k}}^\dagger+h.c.\right)\label{Hright}
\end{eqnarray}
with $\mu_+=v_{F+}k_{F+}$ and $v_{F+}$ is the Fermi velocity at this Fermi surface. $h.c.$ stands for Hermitian conjugate of the pairing term. It should be noticed that a uniform pairing is used here to reproduce the pairing near the Fermi surface in the model (\ref{Hminimal}), which is possible because the structure of the pairing is determined by time-reversal symmetry, such that the topologically trivial and nontrivial pairings have the same form when restricted to one Fermi surface. The main difference between topologically trivial and nontrivial pairings is the relative sign between the pairing order parameters at the two Fermi surfaces. The other Fermi surface with spin helicity $-1$ can be represented similarly by a Weyl Fermion with opposite helicity:
\begin{eqnarray}
H_L&=&\sum_{\bf k}\psi_{L{\bf k}}^\dagger\left[-v_{F-}\sigma\cdot{\bf k}-\mu_-\right]\psi_{L{\bf k}}\nonumber\\
& &-\frac12\left(\sum_{\bf k}\Delta_0k_{F+}\psi_{L-{\bf k}}^\dagger i\sigma_y\psi_{L{\bf k}}^\dagger+h.c.\right)\label{Hleft}
\end{eqnarray}

The Hamiltonian $H_L+H_R$ correctly reproduces the low energy behavior of the minimal model (\ref{Hminimal}) in the weak pairing limit. % $\Delta_0k_{F\pm}\ll v_{F\pm}\left(k_{F-}-k_{F+}\right)$, or equivalently $\Delta_0\ll \alpha$.
More generically, one can represent each Fermi surface with Chern number $1$ ($-1$) by a Weyl fermion with right-hand (left-hand) helicity. The Fermi surfaces with higher Chern number can be similarly represented by multiple channels of Weyl fermions.

The advantage of the Weyl Fermion representation is its relation to the $(3+1)$-d topological superconductor to $(4+1)$-d topological insulators. As is proved in Ref. [\onlinecite{qi2008b}], time-reversal invariant $(4+1)$-d insulators are classified by an integer valued topological invariant, the second Chern number
\begin{eqnarray}
C_2=\frac1{32\pi^2}\int d^4{\bf k}\epsilon^{ijkl}{\rm Tr}\left[f_{ij}f_{kl}\right]
\end{eqnarray}
with $f_{ij}=\partial_ia_j-\partial_ja_i+i\left[a_i,a_j\right]$ and $a_i^{nm}=-i\bra{n{\bf k}}\partial_{k_i}\ket{m{\bf k}}$ the $U(N)$ Berry phase gauge field for a system with $N$ bands occupied. $\ket{n{\bf k}}$ labels the occupied bands. Coupled with a charge $U(1)$ gauge field $A_\mu$, the system with Chern number $C_2$ has a topological response described by the $(4+1)$-d Chern-Simons term
\begin{eqnarray}
S_{\rm CS}\left[A_a\right]=\frac{C_2}{24\pi^2}\int d^5x\epsilon^{abcde}A_a\partial_b A_c\partial_d A_e\label{CS4d}
\end{eqnarray}
with $a,b,..$ denoting the $(4+1)$-d space-time indices $0,1,2,3,4$. In a geometry with open boundary, the $(4+1)$-dimensional topological insulators have surface states on the $(3+1)$-d boundary described by Weyl fermions. The net helicity of Weyl fermions is given by $N_+-N_-=C_2$, with $N_\pm$ the number of right-handed/left-handed Weyl fermions. The charge transport determined by the Chern-Simons term (\ref{CS4d})
\begin{eqnarray}
j_a=\frac1{32\pi^2}\epsilon^{abcde}F_{bc}F_{de}
\end{eqnarray}
correctly reproduces the axial anomaly of the Weyl fermions on the boundary\cite{callan1985}.

\begin{figure}
\includegraphics[width=5.5cm, height=3.5cm]{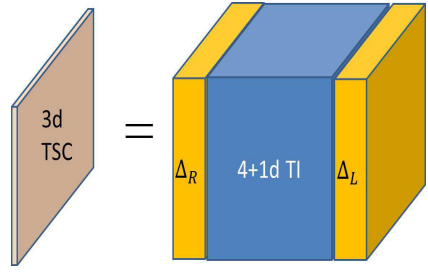}
\caption{Representation of the 3+1d topological superconductor on the spatial manifold $M_3$ as the boundary state of
a 4+1d topological insulator defined on $M_3\times I$.}
\label{fig1}
\end{figure}

Now consider a $(4+1)$-d topological insulator with Chern number $C_2=1$ on a slab geometry $T^3\times I$ (with $T^3$ the torus and $I$ the interval), as shown in Fig. \ref{fig1} (a). For example a representative Hamiltonian in this class is the lattice model introduced in Ref. [\onlinecite{qi2008b}]:
\begin{eqnarray}
H_{\rm bulk}=\sum_{\bf k}c_{\bf k}^\dagger \left[\sum_{i=1}^4\sin k_i \Gamma^i+\left(m+\sum_i\cos k_i\right)\Gamma^0\right]c_{\bf k}\nonumber\\
\label{TI4d}
\end{eqnarray}
with $\Gamma^a,~a=0,1,..,4$ Hermitian matrices satisfying the Clifford algebra $\left\{\Gamma^a,\Gamma^b\right\}=2\delta^{ab}$. For $-4<m<-2$ the system is a topological insulator with Chern number $C_2=1$, and the surface state on each boundary of $T^3\times I$ consists of a single copy of Weyl fermion, with opposite helicity on the two boundaries. The system is time-reversal invariant, with the time-reversal transformation matrix $\mathcal{T}$ satisfying $\mathcal{T}^{-1}\Gamma_i^*\mathcal{T}=-\Gamma_i,~i=1,2,3,4$, $\mathcal{T}^{-1}\Gamma_0^*\mathcal{T}=\Gamma_0$.

We denote the periodic dimensions as $x_{1,2,3}\in[0,L_{1,2,3}]$ and the dimension with open boundary conditions as $x_4\in[0,L_4]$. In the limit of finite $L_4$ and $L_{1,2,3}\rightarrow \infty$, the $(4+1)$-d system can be considered as a $(3+1)$-d system with periodic boundary conditions, and a large number ($4L_4$) of bands per site. However, the low energy effective theory of this system only consists of two massless Weyl fermions with opposite helicity.

Now we relate this model to the $(3+1)$-d topological superconductor by considering a superconducting pairing term at each boundary of a$(4+1)$-d system. Physically, one can consider the geometry shown in Fig. (\ref{fig1}) (b), with the topological insulator sandwiched between two thin layers of $(3+1)$-dimensional ordinary $s$-wave superconductors. Denoting the phases of the two superconductors by $\theta_L$ and $\theta_R$, the low energy effective theory of this sandwich structure is given by
\begin{eqnarray}
H&=&\sum_{\bf k}v_F\left[\psi^\dagger_{R{\bf k}}\sigma\cdot {\bf k}\psi_{R{\bf k}}-\psi_{L{\bf k}}^\dagger \sigma\cdot{\bf k}\psi_{L{\bf k}}\right]\nonumber\\
& &+\frac12\sum_{\bf k}\left[\left|\Delta_R\right|e^{i\theta_R}\psi_{R-{\bf k}}^\dagger i\sigma_y\psi_{R{\bf k}}^\dagger\right.\nonumber\\
& &\left.+\left|\Delta_L\right|e^{i\theta_L}\psi_{L-{\bf k}}^\dagger i\sigma_y\psi_{L{\bf k}}^\dagger+h.c.\right]
\end{eqnarray}
Compare this effective theory with Eq. (\ref{Hright}) and (\ref{Hleft}), one can see clearly that the SC-TI-SC sandwich structure is topologically equivalent to a $(3+1)$-d topological superconductor described by the Hamiltonian (\ref{Hminimal}) if $\theta_L=0,~\theta_R=\pi$.

In a similar way, the topological superconductor with higher winding number $N$ can also be represented by a similar $(4+1)$-d ``sandwich" model with the bulk Chern number $C_2=N$, and two surface superconductors with opposite sign of pairing. The surface states on each boundary consist of $N$ Weyl fermions, each of which has a Fermi surface with first Chern number $C_1=1$. If the two surfaces have opposite sign of pairing, one can verify that the topological invariant of the 3d TSC is $N$.

\subsection{Axion field theory}

The mapping of the $(3+1)$-d TSC to the $(4+1)$-d model discussed above may seem uneconomic at first sight, since the $(4+1)$-d model contains many more high energy degrees of freedom. However, the $(4+1)$-d model has the advantage of spatially separating the left- and right-handed Weyl fermions to different surfaces, which is convenient for deriving the effective field theory description of the TSC. To obtain the effective field theory, we consider a uniform superconducting pairing term on the boundaries
\begin{eqnarray}
H_{\rm SC}&=&\sum_{x_\mu}\Delta_0e^{-i\theta_L(x_\mu)} c^T(x_\mu,x_4=0)\mathcal{T}c(x_\mu,x_4=0)+h.c. \nonumber\\
& &+\sum_{x_\mu}\Delta_0e^{-i\theta_R(x_\mu)} c^T(x_\mu,x_4=L_4)\mathcal{T}c(x_\mu,x_4=L_4)\nonumber\\
& &+h.c.\label{Hsc}
\end{eqnarray}
Here the two terms are pairing of electrons on the left and right boundary defined by the $4$-th coordinate $x_4=0$ and $L_4$ respectively. We consider the pairing amplitude $\Delta_0$ as a constant, and keep the phase fluctuation $\theta_{R(L)}(x_\mu)$ as generic fields depending on the boundary $(3+1)$-d space-time coordinates $x_\mu,~\mu=0,1,2,3$. The physical electromagnetic field in the $(3+1)$-d space-time is represented by a $U(1)$ gauge field defined in the $(4+1)$-d space-time, with the restriction that $A_4\equiv 0$ and $A_\mu=A_\mu(x_\nu)$ only depends on the $(3+1)$-d coordinates $x_\nu$. The fermions are minimally coupled to the gauge field. %In the following we consider the amplitude of the pairing field $\left|\Delta_{R}\right|=\left|\Delta_L\right|=\Delta_0$ as a constant, and only keep the phase fluctuation $\theta_{R(L)}(x_\mu)$ and the gauge field $A_\mu$.
The effective action of the boson fields $\theta_R,\theta_L,A_\mu$ is defined by integrating out the fermion in the action:
\begin{widetext}
\begin{eqnarray}
\exp\left(iS_{\rm eff}\left[A_\mu,\theta_L,\theta_R\right]\right)&\equiv&\int D\bar{c} Dc \exp\left(iS\left[\bar{c},c,A_\mu,\theta_L,\theta_R\right]\right)\nonumber\\
S\left[\bar{c},c,A_\mu,\theta_L,\theta_R\right]&=&\int dt\left[\sum_i\bar{c}_i\left(\partial_t-iA_0\right)c_i-H_{\rm bulk}\left[\bar{c},c,A_i\right]-H_{\rm SC}\left[\bar{c},c,\theta_L,\theta_R\right]\right]\label{Seff}
\end{eqnarray}
\end{widetext}
with $H_{\rm bulk}\left[\bar{c},c,A_i\right]$ defined by the Hamiltonian of topological insulator given in Eq. (\ref{TI4d}), and $H_{\rm SC}$ the boundary pairing terms (\ref{Hsc}).

To compute the effective action it is helpful to notice the gauge invariance of the action $S\left[\bar{c},c,A_\mu,\theta_L,\theta_R\right]$ under the following gauge transformation:
\begin{eqnarray}
c(x_a)&\rightarrow &c(x_a)e^{i\varphi(x_a)}\nonumber\\
A_a(x_b)&\rightarrow&A_a(x_b)+\partial_a\varphi\nonumber\\
\theta_R(x_\mu)&\rightarrow &\theta_R(x_\mu)+2\varphi(x_\mu,x_4=0)\nonumber\\
\theta_L(x_\mu)&\rightarrow &\theta_L(x_\mu)+2\varphi(x_\mu,x_4=L_4)
\end{eqnarray}
Here $x_a,a=0,1,2,3,4$ denotes the $(4+1)$-d coordinates. In particular if we make the special choice
\begin{eqnarray}
\varphi(x_a)=-\frac1{2L_4}\left.[\theta_L(L_4-x_4)+\theta_R x_4\right]
\end{eqnarray}
$\theta_L,\theta_R$ are canceled by the gauge transformation and the gauge field is transformed to $\tilde{A}_a=A_a+\partial_a\varphi$. %The action is transformed to $S\left[\bar{c},c,A_a+\partial_a\varphi,0,0\right]$.
This way, the dependence of the effective action on the pairing phases $\theta_L,\theta_R$ has been absorbed into the dependence to the $U(1)$ gauge field, and the $U(1)$ gauge field now obtains the $4$-th component $\tilde{A}_4=\partial_4\varphi=\left(\theta_L-\theta_R\right)/2L_4$.

Since the pairing terms are only introduced on the boundary, the bulk remains a $(4+1)$-d topological insulator coupled with the gauge field $\tilde{A}_a$. Consequently, the effective action of the gauge field still contains the the Chern-Simons term given in Eq. (\ref{CS4d}) with coefficient $C_2=1$. Written in the physical degrees of freedom $\theta_{R(L)}$ and $A_\mu$, we have
\begin{eqnarray}
S_{\rm eff}&=&S_{\rm CS}\left[\tilde{A}_a\right]=\frac1{24\pi^2}\int d^5x\epsilon^{abcde}\tilde{A}_a \partial_b \tilde{A}_c \partial_d \tilde{A}_e\nonumber\\
&=&\frac1{32\pi^2}\int d^4x\epsilon^{\mu\nu\sigma\tau}\frac{\theta_L-\theta_R}2F_{\mu\nu}F_{\sigma\tau}.\label{Saxion}
\end{eqnarray}
Thus we see that the Chern-Simons term on the slab geometry is reduced to the $(3+1)$-d axion term with the $U(1)$ field $\theta_L-\theta_R$ coupled to the gauge field in an axionic coupling. Besides the topological term (\ref{Saxion}), we also have the ordinary terms in a superconductor, including Higgs term and the higher order terms such as the Maxwell term. Since the two boundaries represent the two Fermi surfaces in the physical superconductor, the two superconducting phases $\theta_L,~\theta_R$ are generically not decoupled since there is only a global $U(1)$ symmetry. Thus Josephson type coupling terms $\cos(\theta_L-\theta_R)$ are also allowed. With these considerations, we finally obtain the following effective action of the $(3+1)$-d topological superconductor:
\begin{widetext}
\begin{eqnarray}
S_{\rm eff}&=&\int d^4x\left[\frac{\theta_L-\theta_R}{64\pi^2}\epsilon^{\mu\nu\sigma\tau} F_{\mu\nu}F_{\sigma\tau}
-\frac1{4e^2}F_{\mu\nu}F^{\mu\nu}+\frac12\rho_L\left(\partial_\mu \theta_L-2A_\mu\right)^2+\frac12\rho_R\left(\partial_\mu \theta_R-2A_\mu\right)^2+J\cos\left(\theta_L-\theta_R\right)\right].\nonumber\\\label{SeffTSC}
\end{eqnarray}
\end{widetext}

The effective action (\ref{SeffTSC}) is the central result of this work. It
describes the response of the fermions to superconducting phase fluctuations $\theta_{R},\theta_L$ of the two Fermi surfaces, and to the electromagnetic field. It should be noticed that the effective action applies to both topological and trivial superconductors depending on the ground state values of $\theta_L,\theta_R$. As has been discussed in Sec. II, $\theta_L=0,~\theta_R=\pi$ corresponds to a topological superconductor, and $\theta_L=\theta_R=0$ corresponds to a trivial superconductor. In the effective action (\ref{SeffTSC}), the ground state value of $\theta_L-\theta_R$ can be controlled by the sign of $J$. $J>0$ ($J<0$) describes topological (trivial) superconductor. { Since this effective theory is obtained by integrating out fermions, it only applies to the energy scale below the BCS gap $E<|\Delta_L|,~|\Delta_R|$. Therefore the coupling between the two Fermi surfaces must be weak enough such that the mass of the mode $\theta_L-\theta_R$ $m^2\equiv 2J/\rho\ll |\Delta_{L,R}|^2$, in order for the effective theory to be meaningful.}
%XLnote: the paragraph above is modified.

The approach of $(4+1)$-d regularization can be generalized to generic TSC with multiple Fermi surfaces. In the weak pairing limit, a superconducting phase $\theta_i$ can be defined on each Fermi surface. If a Fermi surface has the Chern number $C_{1i}$, it is equivalent to $C_{1i}$ copies of the Weyl fermion studied above. The axion term in the electromagnetic response is given by a sum over all Fermi surfaces. Thus we obtain the following generic effective action:

\begin{widetext}
\begin{eqnarray}
S_{\rm eff}=\int d^4x\left[\frac{1}{64\pi^2}\sum_iC_{1i}\theta_i\epsilon^{\mu\nu\sigma\tau} F_{\mu\nu}F_{\sigma\tau}
-\frac1{4e^2}F_{\mu\nu}F^{\mu\nu}+\frac12\sum_i\rho_i\left(\partial_\mu \theta_i-2A_\mu\right)^2+\sum_{i<j}J_{ij}\cos\left(\theta_i-\theta_j\right)\right]\label{SeffTSC2}
\end{eqnarray}
\end{widetext}
It is worthwhile to note that  a time-reversal transformation acts by $\theta_i\rightarrow -\theta_i,~A_0\rightarrow A_0,~A_{x,y,z}\rightarrow -A_{x,y,z}$, and the effective action is invariant, as expected. The Josephson coupling between the superconducting phases of different Fermi surfaces is always in the form of $\cos(\theta_i-\theta_j)$ as is required by time-reversal symmetry. The ground state value of $\theta_i$ (up to an overall constant) is determined by the Josephson couplings $J_{ij}$. In general, it is possible that the coupling is frustrated and the ground state spontaneously breaks time-reversal symmetry, although we are mainly interested in the situation with time-reversal symmetry preserved, with $\theta_i=0$ or $\pi$ for each $i$. In that case the topological invariant (\ref{TopoInv}) is given by
\begin{eqnarray}
N=\frac12\sum_{i}C_{1i}e^{i\theta_i}\label{TopoInv2}
\end{eqnarray}

To summarize this section, we have obtained the generic effective action (\ref{SeffTSC2}) of $(3+1)$-d superconductors coupled with the electromagnetic field and  superconducting phase fluctuations. %which are the central results of this work.
The most important term in the action is the axionic coupling between the superconducting phase fluctuations and the electromagnetic field, which is a consequence of the nontrivial Chern number of the Fermi surface. The effective action describes both topological trivial and nontrivial superconductors. Although not explicit in the effective action, the topological invariant $N$ is determined by the Chern numbers $C_{1i}$ and the ground state value of $\theta_i$, which is thus determined for a given effective action (\ref{SeffTSC2}).

\section{Topological effects described by the axion field theory}\label{sec:consequence}

One of the main advantages of an effective field theory description is that it directly describes observable topological response properties of the topological state of matter. Having obtained the effective field theory (\ref{SeffTSC2}), we now discuss the topological effects described by this theory.

\subsection{Anomaly and chiral vortex lines}

Naively, the effective action (\ref{SeffTSC2}) is dominated by the Higgs terms $\left(\partial_\mu \theta_i-2A_\mu\right)^2$, which leads to Meissner effect and screens the electromagnetic field in the superconductor. With zero field strength $F_{\mu\nu}$ in the superconductor, the topological term will have no physical consequence. However, the situation becomes nontrivial when vortex lines are considered in a type-II superconductor. Around an ordinary Abrikosov vortex line, the phases $\theta_i$ on all Fermi surfaces have the same winding number, so that the phase combination $\sum_iC_{1i}\theta_i$ in the topological term has no winding. To see the nontrivial consequence of the topological term, it is essential to consider a ``chiral vortex line" where only some of the $\theta_i$'s have a vorticity. In the following we will study the anomaly at presence of such chiral vortex lines, as a consequence of the topological term.

We start by considering the equation of motion of the gauge field $A_\mu$ determined by the action (\ref{SeffTSC2}):
\begin{eqnarray}
2\sum_i\rho_i\left(\partial_\mu\theta_i-2A_\mu\right)&=&\sum_i\frac{C_{1i}}{8\pi^2}\epsilon^{\mu\nu\sigma\tau}\partial_\nu\theta_i
\partial_\sigma A_\tau\label{EoM}
%\nonumber\\
%\rho_i\partial_\mu\left(\partial_\mu\theta_i-2A_\mu\right)&=&\frac{C_{1i}}{64\pi^2}\epsilon^{\mu\nu\sigma\tau}F_{\mu\nu}F_{\sigma\tau}-J\sum_{j\neq i}\sin\left(\theta_i-\theta_j\right)
\end{eqnarray}
The charge current is defined as
\begin{eqnarray}
j_\mu=2\sum_i\rho_i\left(\partial_\mu\theta_i-2A_\mu\right)\label{chargecurrent}
\end{eqnarray}
which should be conserved due to the global $U(1)$ symmetry. However, with the $F\wedge F$ term in the action (\ref{SeffTSC2}), there is an anomaly in  charge conservation. According to Eqs. (\ref{EoM}) and (\ref{chargecurrent}), we obtain
\begin{eqnarray}
\partial^\mu j_\mu&=&\sum_i\frac{C_{1i}}{8\pi^2}\epsilon^{\mu\nu\sigma\tau}\partial_\mu\partial_\nu\theta_i\partial_\sigma A_\tau\nonumber\\
&\equiv&\frac1{4\pi}\sum_iC_{1i}J_{{\rm v}i}^{\mu\nu}\partial_\mu A_\nu
\end{eqnarray}
with
\begin{eqnarray}
J_{{\rm v}i}^{\mu\nu}&=&\frac1{2\pi}\epsilon^{\mu\nu\sigma\tau}\partial_\sigma\partial_\tau\theta_i
\end{eqnarray}
the antisymmetric tensor field for the vortex current. For example, if there is only a vortex line for $\theta_1$ along the $z$ direction at coordinate $x=y=0$, with the Chern number $C_{11}=1$, we have
\begin{eqnarray}
J_{{\rm v}1}^{zt}=-J_{{\rm v}1}^{tz}=\delta(x)\delta(y)
\end{eqnarray}
with other components vanishing. The anomaly is given by
\begin{eqnarray}
\int d^4x\partial^\mu j_\mu=\frac1{4\pi}\int_{\rm vortex} dzdt F_{zt}\label{MajoranaAnomaly}
\end{eqnarray}
which is half of the Chern number of the electric field $F_{\mu\nu}$ in the $zt$ plane. Physically, such an anomaly is related to the fact that there is a $1+1$-D Majorana-Weyl fermion propagating along the vortex line\cite{qi2009b}. The anomalous term above obtained from the effective theory cancels the anomaly of the Majorana-Weyl fermion. By comparison, in the more familiar case of a $1+1$d Weyl fermion coupled to gauge field $A_\mu$, the anomaly is given by
\begin{eqnarray}
\partial^aj_a=\frac1{4\pi}\epsilon^{ab}F_{ab}
\end{eqnarray}
with $a,b=0,1$.
For an electric field with flux $2\pi$, $\int d^2x\partial^aj_a=1$ which means that one additional Weyl fermion appears in the $1+1$-d system. Physically, one can view the Weyl fermion as an edge state of the quantum Hall system, and the anomaly is interpreted as a charge pumping from the bulk to the boundary due to the Hall current.\cite{callan1985} Since a Majorana-Weyl fermion carries half the degree of freedom of a Weyl fermion, the anomaly (\ref{MajoranaAnomaly}) is also given by half of that of the Weyl fermion. %{\bf I am confused by the normalization factor, the two equations above have the same coefficients?}

\begin{figure}
\includegraphics[width=7cm]{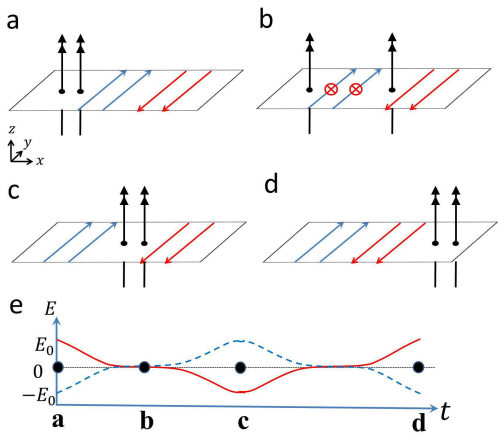}
\caption{(a) A system with two vertical vortex lines (lines with double arrow) and four horizontal chiral vortex lines (lines with single arrow). The two blue (red) horizontal lines are chiral vortices of $\theta_L$ ($\theta_R$), respectively (see text). (b) When one vertical vortex lines move across the two horizontal chiral vortex lines, two Majorana fermion zero modes are created along the horizontal vortex lines. (c) When both vertical vortex lines cross the chiral vortex lines, the Majorana fermions disappear. (d) The same process happens to the right-handed chiral vortex lines when the two vertical vortex lines are moved across them. Due to periodic boundary condition, the configuration in (a) and (d) are considered as the same. In term of energy spectrum, this process corresponds to the level crossing illustrated in (e), with a,b,c,d marking the time corresponding to the configurations showing in (a)-(d). } \label{fig:linking}
\end{figure}

To obtain a more physical understanding of the anomaly equation, the motion of multiple vortex lines needs to be considered, since the electricmagnetic field inducing the anomaly is only present in the superconductor at vortex lines. For simplicity, consider the $N=1$ topological superconductor with two Fermi surfaces satisfying $\theta_L-\theta_R=\pi$ as an example. According to the Higgs term, away from the vortex lines we have
\begin{eqnarray}
\rho_L\left(\partial_\mu \theta_L-2A_\mu\right)+\rho_R\left(\partial_\mu\theta_R-2A_\mu\right)&=&0
%\partial_\mu\left(\theta_L+\theta_R\right)-4A_\mu=0
\end{eqnarray}
so that around a chiral vortex line of $\theta_L$, the flux is
\begin{eqnarray}
\oint {\bf A}\cdot d{\bf l}&=&\frac1{2\left(\rho_L+\rho_R\right)}\oint \left(\rho_L\nabla\theta_L+\rho_R\nabla\theta_R\right)\cdot d{\bf l}\nonumber\\
&=&\frac{\rho_L}{\rho_L+\rho_R}\pi
\end{eqnarray}
in which the loop integral is taken around a loop enclosing the chiral vortex line. {Therefore the chiral vortex line carries a flux of $\frac{\rho_L}{\rho_L+\rho_R}\frac{hc}{4e}$ in Gauss units. Similarly, the chiral vortex line of $\theta_R$ carries a flux of $\frac{\rho_R}{\rho_L+\rho_R}\frac{hc}{4e}$. } %XLnote: This sentence and the equation above are modified.
Consider the configuration illustrated in Fig. \ref{fig:linking} with $4$ chiral vortex lines, $2$ for each of $\theta_L$ and $\theta_R$, along the $y$ direction, and $2$ ordinary vortex lines along the $z$ direction. Periodic boundary conditions are assumed along all three directions $x,y,z$. It should be noted that in such a compact manifold, the flux of the gauge field $F_{\mu\nu}$ in each plane $xy,yz,zx$ must be quantized in unit of $2\pi$, which is why the number of chiral vortex lines must be a multiple of $4$, and that of ordinary vortex lines must be even.

Now consider the motion of $z$ vortex lines across the chiral ones, as is illustrated in Fig. \ref{fig:linking}. When both of the $z$ vortex lines move from the left to the right of the chiral ones, due to periodic boundary conditions, they can return to their original position, so that such a process is a periodic evolution of the system. In the following we will show that such a vortex motion corresponds to an instanton process of $F_{\mu\nu}$ with nontrivial $\int d^4x \epsilon^{\mu\nu\sigma\tau}F_{\mu\nu}F_{\sigma\tau}$.

Since the field strength is only nonzero in the vortex cores, the term $\epsilon^{\mu\nu\sigma\tau}F_{\mu\nu}F_{\sigma\tau}$ vanishes except when two vortex lines cross each other. The field strength corresponds to a static $y$ chiral vortex line at position $(x_0,z_0)$ is
\begin{eqnarray}
F_{zx}^{(1)}=\frac{\pi}2\delta(x-x_0)\delta(z-z_0)
\end{eqnarray}
with other components vanishing. Similarly for an ordinary vortex line along the $z$ direction with coordinates $(x_0(t)=vt,y_0)$, the field strength is
\begin{eqnarray}
F_{xy}^{(2)}&=&\pi \delta(x-vt)\delta(y-y_0)\nonumber\\
F_{ty}^{(2)}&=&-v\pi \delta(x-vt)\delta(y-y_0)
\end{eqnarray}
Therefore the contribution of the two vortex lines to the topological term is
\begin{eqnarray}
\epsilon^{\mu\nu\sigma\tau}F_{\mu\nu}F_{\sigma\tau}&=&2\epsilon^{\mu\nu\sigma\tau}F^{(1)}_{\mu\nu}F^{(2)}_{\sigma\tau}
=-8F^{(1)}_{zx}F^{(2)}_{ty}\nonumber\\
&=&4{\pi^2}v\delta(x-vt)\delta(y-y_0)\delta(x-x_0)\delta(z-z_0)\nonumber\\
&=&4{\pi^2}\delta(t-\frac{x_0}{v})\delta(y-y_0)\delta(x-x_0)\delta(z-z_0)\nonumber\\
&=&4\pi^2\delta^4(x_\mu-x_{0\mu})
\end{eqnarray}
with $\delta^4$ denoting the $4$-dimensional $\delta$ function and $x_{0\mu}=(x_0/v,x_0,y_0,z_0)$ the crossing point in space-time. In the process shown in Fig. \ref{fig:linking}, there are $8$ such vortex crossing points. If we denote the location of the $8$ crossing points by $x_n^\mu$, $n=1,2,...,8$, the topological term for such a process is
\begin{eqnarray}
S_{\rm topo}&\equiv &\int d^4x\frac{\theta_L-\theta_R}{64\pi^2}\epsilon^{\mu\nu\sigma\tau}F_{\mu\nu}F_{\sigma\tau}\nonumber\\
&=&\frac1{16}\sum_{n=1}^8\left(\theta_{L}(x_n)-\theta_{R}(x_n)\right)
\end{eqnarray}

To understand the implication of such a term, consider the transformation $\theta_L(x)\rightarrow \theta_L(x)+2\pi$ with $\theta_R$ invariant. The change of the topological term is $\Delta S_{\rm topo}=\pi$, so that the partition function changes sign $Z\rightarrow -Z$. This seems to be contradictory with the fact that $\theta_L$ is the phase of the order parameter and is periodic with the period of $2\pi$. However, this is exactly the consequence of the anomaly in the Majorana fermion system. The phase rotation $\theta_L\rightarrow \theta+2\pi$ corresponds to a phase rotation of $\pi$ for electrons $\psi_L(x)\rightarrow -\psi_L(x)$. In such a transformation, a many-body state obtains a phase $(-1)^{N_L}$ with $N_L$ the number of left-handed fermions. $(-1)^{N_L}$ is the fermion number parity of the left-handed fermions. In the adiabatic limit, the propagator $e^{iS_{\rm topo}}$ can be considered as a propagator of the ground state back to itself:
\begin{eqnarray}
e^{iS_{\rm topo}}=\left\langle G\right|e^{-i\int_{-\infty}^{+\infty} dt H(t)}\left|G\right\rangle
\end{eqnarray}
The fact that $e^{iS_{\rm topo}}\rightarrow -e^{iS_{\rm topo}}$ in the global transformation $\psi_L\rightarrow -\psi_L$ can be interpreted as the initial and final states having opposite fermion number parity for the left-handed fermion. Similarly, the initial and final state also have opposite fermion number parity for the right-handed fermion. This effect can be understood as a residual effect of the anomaly in Weyl fermions. If we consider the normal state of the topological superconductor, it is topologically equivalent to a pair of Weyl fermions with opposite chirality. In a background field $F_{\mu\nu}$, the Weyl fermions have the anomaly
\begin{eqnarray}
\partial_\mu j^\mu_L=-\partial_\mu j^\mu_R=\frac1{32\pi^2}\epsilon^{\mu\nu\sigma\tau}F_{\mu\nu}F_{\sigma\tau}
\end{eqnarray}
In an instanton configuration with $C_2=\frac1{32\pi^2}\epsilon^{\mu\nu\sigma\tau}F_{\mu\nu}F_{\sigma\tau}=1$, we have $\int d^4x\partial_\mu j^\mu_L=-\int d^4x\partial_\mu j^\mu_R=1$, which means one left-handed Weyl fermion becomes right-handed after the instanton process. When we consider superconductivity, the fermion number conservation is broken, but the fermion number parity conservation is still preserved. Consequently the anomaly is still well-defined for instanton processes with odd $C_2$. The vortex line crossing process we discussed above is an example of a configuration with $C_2=1$.

More microscopically, such an anomaly can be related to some spectral flow of the Majorana fermions, as is illustrated in Fig. \ref{fig:linking}. This is similar to the spectral flow picture of the $Z_2$ Witten anomaly\cite{witten1982}. Each chiral vortex line hosts a $(1+1)$-d Majorana-Weyl fermion, which can be seen by explicitly solving the Majorana fermion spectrum with a vortex line of the mass term. This is the Majorana version of the axion string. When periodic boundary conditions are considered for the chiral vortex lines along the $y$ direction, whether or not each chiral vortex line has a Majorana zero mode at exactly zero energy depends on the boundary condition of the Majorana-Weyl fermion. We start from the configuration in Fig. \ref{fig:linking} (a) and assume that the horizontal vortex lines have anti-periodic boundary conditions and thus have no zero modes. When the perpendicular vortex line crosses the first horizontal one, the boundary condition for the Majorana-Weyl fermion along the horizontal chiral vortex line is changed, so that a pair of Majorana zero modes appear on the two vortex lines which have just crossed each other. Subsequently, when the perpendicular vortex line crosses the second chiral vortex line, there are two Majorana zero modes on the two horizontal vortex lines. When the second perpendicular vortex line crosses them, the boundary condition is changed back to anti-periodic and thus the zero modes disappear. The spectral flow in the whole process is summarized in Fig. \ref{fig:linking} (e) which contains a level crossing between electron and hole states. Due to this level crossing, one would conclude that the ground state on the left side and that on the right side have opposite fermion number parity. When the perpendicular vortex lines cross the two right-handed chiral vortex lines, a similar level crossing occurs for the right-handed fermions. Therefore the consequence of the instanton process is the pumping of a fermion from left-handed Fermi surface to the right-handed Fermi surface.

\subsection{Chiral vortex lines and $s$-wave proximity effect}

%XLnote: this paragraph is modified.

{Since there is no independent $U(1)$ symmetry for $\theta_L$ and $\theta_R$, one may expect the chiral vortex lines of $\theta_L$ and $\theta_R$ to be confined with each other, which seems to make it difficult to physically realize the vortex motion discussed above. However, in the following we will show that such chiral vortex lines can actually be realized experimentally on the interface between a topological superconductor and a trivial $s$-wave superconductor. By making use of the interface, the chiral vortex lines are deconfined with each other as long as they stay in the interface plane.}

As is discussed earlier, a trivial $s$-wave superconductor with $N=0$ corresponds to setting $\theta_L=\theta_R$. If the phases are $(\theta_L,\theta_R)=(0,0)$ for the trivial superconductor and $(\theta_L,\theta_R)=(0,\pi)$ for the nontrivial one, the interface between trivial and topological superconductors is a $\pi$ phase domain wall of $\theta_R$.
%XLnote: I added two sentences in the following.
{ (Another possibility is to have $(\theta_L,\theta_R)=(\pi,0)$ in the topological superconductor, so that the interface is a $\theta_L$ domain wall. This case can be analyzed in the same way.)
If time-reversal symmetry is preserved, $\theta_R$ has to jump from $0$ to $\pi$ in the interface between the two materials, which means the pairing amplitude $\left|\Delta_R\right|$ must vanish somewhere near the surface.  }
If time-reversal symmetry breaking is allowed on the interface, $\theta_R$ can change from $0$ to $\pi$ smoothly across the surface. Since $\theta_R$ is periodic in $2\pi$, there are different ways to interpolate between $\theta_R=0$ and $\theta_R=\pi$, corresponding to
\begin{eqnarray}
\Delta\theta_R=\int dz\partial_z\theta_R=\pi+2n\pi\nonumber
\end{eqnarray}
A vortex of $\theta_R$ can be created if we consider a domain wall between different interpolations. %As is shown in Fig. \ref{proximity} (a),
 If $\Delta\theta_R=-\pi$ for $x<0$ and $\Delta\theta_R=\pi$ for $x>0$, a chiral vortex line is created at the line $x=0,z=0$.

\begin{figure}
\includegraphics[width=8cm]{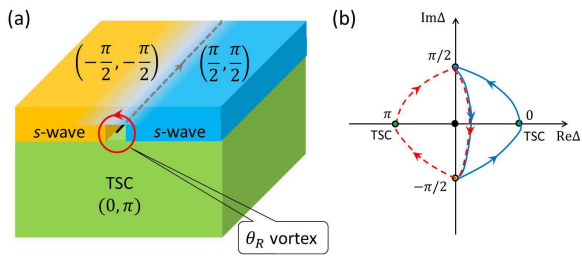}
\caption{(a) Illustration of the tri-junction configuration with two $s$-wave superconductors on top of a topological superconductor (TSC). The phase of the $s$-wave superconductors are $-\pi/2$ in the blue region and $\pi/2$ in the yellow region. The vector $(0,\pi)$, $(\pi/2,\pi/2)$ etc stands for the value of $\left(\theta_L,\theta_R\right)$ in the corresponding regions. (b) The evolution of phases $\theta_L$ (blue solid line) and $\theta_R$ (red dashed line) along the red circle path marked in panel (a). In such a configuration $\theta_R$ has winding number $-1$ while $\theta_L$ has winding number $0$. Depending on the sign of the phase gradient between the two $s$-wave superconductors (the position of the vertical path connecting $\pi/2$ and $-\pi/2$ relative to the origin), one may also have winding number $0$ for $\theta_R$ and $1$ for $\theta_L$. Independent from this choice, there is always a Majorana-Weyl fermion propagating along the junction, as is illustrated by the dashed line in panel (a). %VAC stands for vacuum which is topological equivalent to an $s$-wave superconductor.
} \label{fig:proximity}
\end{figure}

%({\bf The following paragraph and fig 3 are modified.--XL})%XLnote
Such a chiral vortex line can be realized in a tri-junction configuration shown in Fig. \ref{fig:proximity} which consists of two $s$-wave superconductor films on the surface of the 3D topological superconductor.\cite{wang2011} The phases of the superconductors can be tuned such that $(\theta_L,\theta_R)=\pm(\pi/2,\pi/2)$ for the two $s$-wave superconductors, and $(\theta_L,\theta_R)=(0,\pi)$ for the topological superconductor. Across the junctions between the three superconductors, the phases $\theta_L$ and $\theta_R$ interpolate continuously. Therefore if we follow the evolution of the phases along the counterclockwise circle around the junction (shown by the red circle in Fig. \ref{fig:proximity} (a)), the evolution of $\theta_L$ and $\theta_R$ are shown in Fig. \ref{fig:proximity} (b) by the blue and red paths, respectively. From this picture we see immediately that the winding number $N_L$ and $N_R$ of $\theta_L,~\theta_R$ are different and we always have $N_L-N_R=1$. Depending on whether the interpolation from $-\pi/2$ to $+\pi/2$ across the junction between two $s$-wave superconductors is done through $0$ or $\pi$, we may have $(N_L,N_R)=(0,-1)$ or $(1,0)$, both of which correspond to the same Majorana-Weyl fermion propagating along the junction as is illustated in Fig. \ref{fig:proximity} (a) by the dashed line with arrow.

%The vacuum above the surface of TSC is topologically equivalent to a trivial superconductor with $\theta_L=\theta_R=0$. The surface of TSC is in proximity with two $s$-wave superconductor films, which have phases $\theta_L=\theta_R=\pm\pi/2$, respectively. For such a configuration it is natural to expect that the phase interpolates continuously from $0$ to $\pm \pi/2$ to $0$ or $\pi$ across the surface. Therefore it is straightforward to see that $\theta_L$ has no vorticity and $\theta_R$ has a vortex line along the Josephson junction. In other words, such a Josephson junction is a chiral vortex line.

The Majorana-Weyl fermion along the junction in this case can also be understood from an alternative surface state point of view. As has been discussed in Ref. \cite{wang2011}, the $s$-wave superconductor on the surface with phase $\pi/2$ breaks time-reversal symmetry and leads to a mass term of the Majorana fermion on the surface. The surface state of the TSC is described by the low energy effective theory
\begin{eqnarray}
 H=\int d^2x \eta^T\left[\sigma_x(-i\partial_x)+\sigma_z(-i\partial_y)+m(x)\sigma_y\right]\eta
\end{eqnarray}
with $m(x)=m (-m)$ for $x>0$ and $x<0$, respectively. The Josephson junction is a mass domain wall for the surface Majorana fermion, which induces a Majorana-Weyl fermion bound state propagating on it\cite{teo2010,wang2011}.

This construction can be generalized to systems with multiple Fermi surfaces. For a TSC with $\theta_i=0$ or $\pi$ for each Fermi surface, the $s$-wave Josephson junction shown in Fig. \ref{fig:proximity} is a vortex line for each Fermi surface with $\theta_i=\pi$. If the Fermi surface has Chern number $C_{1i}$, the chiral vortex line has $C_{1i}$ Majorana-Weyl fermion modes. (If $C_{1i}<0$, the corresponding Majorana-Weyl fermion has opposite chirality.) Therefore the total number of Majorana-Weyl fermions is
\begin{eqnarray}
N=-\sum_iC_{1i}\frac{1-e^{i\theta_i}}2=\frac12\sum_iC_{1i}e^{i\theta_i}
\end{eqnarray}
which agrees with the bulk topological invariant of the TSC.

%More physical interpretation of such an anomaly can be obtained by considering the motion of multiple vortex lines, which will be discussed in the next subsection.

%explain why this corresponds to Majorana. Two vortex lines correspond to one Weyl fermion.

%\subsection{Linking and crossing of vortex lines}

\section{Conclusion and discussion}\label{sec:summary}

In conclusion, in this work we have derived a new topological field theory description of the three-dimensional TRI TSC. More precisely, our theory describes the dynamics of electromagnetic field coupled with the dynamic phase fluctuation of the superconducting order parameter in a generic $(3+1)$-d superconductor. The topological property of the TSC is determined by the ground state value of the superconducting phases, so that the physical distinction between superconductors with different topological invariants is characterized by our theory. The superconducting phases couple to electromagnetic field in an axionic topological coupling, with the coefficient determined by a topological property---the Chern number of each Fermi surface. As physical consequences of our topological field theory, we discussed the spectral flow and fermion number parity change in the process of mutual linking between chiral vortex lines. Physically, the chiral vortex lines can be realized by Josephson junctions between $s$-wave superconductor films on top of a 3d TRI TSC, so that the prediction of our theory can be verified in future experiments for possible candidate materials of 3d TSC.

Following the same derivation as that in Sec. \ref{sec:axion}, one can also investigate the response to gravitational field which was discussed in Refs. \cite{wang2011,ryu2012b}. If we consider the coupling of fermions to gravitational field in the $(4+1)$-d theory (\ref{Seff}), beside the Chern-Simons term (\ref{Saxion}) one will obtain another mixed Chern-Simons term
\begin{eqnarray}
S_{\rm grav}=\frac{1}{768\pi^2}\int d^5x\epsilon^{abcde}A_aR^f_{gbc}R^g_{fde}\label{GravityTFT4d}
\end{eqnarray}
with $R^a_{bcd}$ the Riemann curvature. If we only consider the gravitational fluctuation of the physical $(3+1)$-d manifold and keep the extra dimension flat, $R^a_{bcd}$ is nonvanished only if no index is in the extra dimension. Therefore this topological term is reduced to
\begin{eqnarray}
S_{\rm grav}=\frac{\theta_L-\theta_R}{1536\pi^2}\int d^5x\epsilon^{\mu\nu\sigma\tau}R^\alpha_{\beta\mu\nu}R^\beta_{\alpha\sigma\tau}\label{GravityTFT}
\end{eqnarray}
 This action can be straightforwardly generalized to the more generic form in analogy to Eq. (\ref{SeffTSC2}):
\begin{eqnarray}
S_{\rm grav}=\frac{1}{1536\pi^2}\sum_iC_{1i}\theta_i\int d^5x\epsilon^{\mu\nu\sigma\tau}R^\alpha_{\beta\mu\nu}R^\beta_{\alpha\sigma\tau}\label{GravityTFTgeneral}
\end{eqnarray}
This result is consistent with the results of Ref. \cite{wang2011,ryu2012b}. However, the apparent contradiction between the integer-valued topological invariant of the bulk TSC and the $Z_2$ value of the coefficient $\theta_L-\theta_R=0,\pi$ in the presence of time-reversal symmetry is resolved in the new derivation, since $\theta_L,\theta_R$ are now interpreted as superconducting phase fields rather than a topological order parameter. The bulk topological invariant is determined by the integer coefficients $C_{1i}$ of this coupling together with the ground state value of the superconducting phase, as is given in Eq. (\ref{TopoInv2}). The main difference between the gravitational and electromagnetic response is that there is no Higgs term in the gravitational response, so that the gravitational field still propagates in the superconductor (if there is an Einstein-Hilbert term in the action). Along a chiral vortex line of $\theta_i$, a gravitational anomaly\cite{alvarez-gaume1985} emerges which is consistent with the presence of a Majorana-Weyl fermion mode along it.

This work is supported by Packard Foundation (XLQ) and the NSF under grant numbers PHY-0969448 (EW) and DMR-0904264 (SCZ).

\bibliography{TI}

\begin{thebibliography}{40}
\expandafter\ifx\csname natexlab\endcsname\relax\def\natexlab#1{#1}\fi
\expandafter\ifx\csname bibnamefont\endcsname\relax
  \def\bibnamefont#1{#1}\fi
\expandafter\ifx\csname bibfnamefont\endcsname\relax
  \def\bibfnamefont#1{#1}\fi
\expandafter\ifx\csname citenamefont\endcsname\relax
  \def\citenamefont#1{#1}\fi
\expandafter\ifx\csname url\endcsname\relax
  \def\url#1{\texttt{#1}}\fi
\expandafter\ifx\csname urlprefix\endcsname\relax\def\urlprefix{URL }\fi
\providecommand{\bibinfo}[2]{#2}
\providecommand{\eprint}[2][]{\url{#2}}

\bibitem[{\citenamefont{von Klitzing et~al.}(1980)\citenamefont{von Klitzing,
  Dorda, and Pepper}}]{klitzing1980}
\bibinfo{author}{\bibfnamefont{K.}~\bibnamefont{von Klitzing}},
  \bibinfo{author}{\bibfnamefont{G.}~\bibnamefont{Dorda}}, \bibnamefont{and}
  \bibinfo{author}{\bibfnamefont{M.}~\bibnamefont{Pepper}},
  \bibinfo{journal}{Phys. Rev. Lett.} \textbf{\bibinfo{volume}{45}},
  \bibinfo{pages}{494} (\bibinfo{year}{1980}).

\bibitem[{\citenamefont{Tsui et~al.}(1982)\citenamefont{Tsui, Stormer, and
  Gossard}}]{tsui1982}
\bibinfo{author}{\bibfnamefont{D.~C.} \bibnamefont{Tsui}},
  \bibinfo{author}{\bibfnamefont{H.~L.} \bibnamefont{Stormer}},
  \bibnamefont{and} \bibinfo{author}{\bibfnamefont{A.~C.}
  \bibnamefont{Gossard}}, \bibinfo{journal}{Phys. Rev. Lett.}
  \textbf{\bibinfo{volume}{48}}, \bibinfo{pages}{1559} (\bibinfo{year}{1982}).

\bibitem[{\citenamefont{Qi and Zhang}(2011)}]{qi2011rmp}
\bibinfo{author}{\bibfnamefont{X.-L.} \bibnamefont{Qi}} \bibnamefont{and}
  \bibinfo{author}{\bibfnamefont{S.-C.} \bibnamefont{Zhang}},
  \bibinfo{journal}{Rev. Mod. Phys.} \textbf{\bibinfo{volume}{83}},
  \bibinfo{pages}{1057} (\bibinfo{year}{2011}).

\bibitem[{\citenamefont{Hasan and Kane}(2010)}]{hasan2010}
\bibinfo{author}{\bibfnamefont{M.~Z.} \bibnamefont{Hasan}} \bibnamefont{and}
  \bibinfo{author}{\bibfnamefont{C.~L.} \bibnamefont{Kane}},
  \bibinfo{journal}{Rev. Mod. Phys.} \textbf{\bibinfo{volume}{82}},
  \bibinfo{pages}{3045} (\bibinfo{year}{2010}).

\bibitem[{\citenamefont{Moore}(2010)}]{moore2010}
\bibinfo{author}{\bibfnamefont{J.~E.} \bibnamefont{Moore}},
  \bibinfo{journal}{Nature} \textbf{\bibinfo{volume}{464}},
  \bibinfo{pages}{194} (\bibinfo{year}{2010}).

\bibitem[{\citenamefont{Read and Green}(2000)}]{read2000}
\bibinfo{author}{\bibfnamefont{N.}~\bibnamefont{Read}} \bibnamefont{and}
  \bibinfo{author}{\bibfnamefont{D.}~\bibnamefont{Green}},
  \bibinfo{journal}{Phys. Rev. B} \textbf{\bibinfo{volume}{61}},
  \bibinfo{pages}{10267} (\bibinfo{year}{2000}).

\bibitem[{\citenamefont{Schnyder et~al.}(2008)\citenamefont{Schnyder, Ryu,
  Furusaki, and Ludwig}}]{schnyder2008}
\bibinfo{author}{\bibfnamefont{A.~P.} \bibnamefont{Schnyder}},
  \bibinfo{author}{\bibfnamefont{S.}~\bibnamefont{Ryu}},
  \bibinfo{author}{\bibfnamefont{A.}~\bibnamefont{Furusaki}}, \bibnamefont{and}
  \bibinfo{author}{\bibfnamefont{A.~W.~W.} \bibnamefont{Ludwig}},
  \bibinfo{journal}{Phys. Rev. B} \textbf{\bibinfo{volume}{78}},
  \bibinfo{pages}{195125} (\bibinfo{year}{2008}).

\bibitem[{\citenamefont{Roy}(2008)}]{roy2008}
\bibinfo{author}{\bibfnamefont{R.}~\bibnamefont{Roy}},
  \bibinfo{howpublished}{e-print arXiv:0803.2868} (\bibinfo{year}{2008}).

\bibitem[{\citenamefont{Qi et~al.}(2009{\natexlab{a}})\citenamefont{Qi, Hughes,
  Raghu, and Zhang}}]{qi2009b}
\bibinfo{author}{\bibfnamefont{X.-L.} \bibnamefont{Qi}},
  \bibinfo{author}{\bibfnamefont{T.~L.} \bibnamefont{Hughes}},
  \bibinfo{author}{\bibfnamefont{S.}~\bibnamefont{Raghu}}, \bibnamefont{and}
  \bibinfo{author}{\bibfnamefont{S.-C.} \bibnamefont{Zhang}},
  \bibinfo{journal}{Phys. Rev. Lett.} \textbf{\bibinfo{volume}{102}},
  \bibinfo{pages}{187001} (\bibinfo{year}{2009}{\natexlab{a}}).

\bibitem[{\citenamefont{Chung and Zhang}(2009)}]{chung2009}
\bibinfo{author}{\bibfnamefont{S.~B.} \bibnamefont{Chung}} \bibnamefont{and}
  \bibinfo{author}{\bibfnamefont{S.~C.} \bibnamefont{Zhang}},
  \bibinfo{journal}{Phys. Rev. Lett.} \textbf{\bibinfo{volume}{103}},
  \bibinfo{pages}{235301} (\bibinfo{year}{2009}).

\bibitem[{\citenamefont{Yan et~al.}(2010)\citenamefont{Yan, Liu, Zhang, Yam,
  Qi, Frauenheim, and Zhang}}]{yan2010a}
\bibinfo{author}{\bibfnamefont{B.}~\bibnamefont{Yan}},
  \bibinfo{author}{\bibfnamefont{C.-X.} \bibnamefont{Liu}},
  \bibinfo{author}{\bibfnamefont{H.}~\bibnamefont{Zhang}},
  \bibinfo{author}{\bibfnamefont{C.~Y.} \bibnamefont{Yam}},
  \bibinfo{author}{\bibfnamefont{X.~L.} \bibnamefont{Qi}},
  \bibinfo{author}{\bibfnamefont{T.}~\bibnamefont{Frauenheim}},
  \bibnamefont{and} \bibinfo{author}{\bibfnamefont{S.~C.} \bibnamefont{Zhang}},
  \bibinfo{journal}{Europhys. Lett.} \textbf{\bibinfo{volume}{90}},
  \bibinfo{pages}{37002} (\bibinfo{year}{2010}).

\bibitem[{\citenamefont{Fu and Berg}(2010)}]{fu2010b}
\bibinfo{author}{\bibfnamefont{L.}~\bibnamefont{Fu}} \bibnamefont{and}
  \bibinfo{author}{\bibfnamefont{E.}~\bibnamefont{Berg}},
  \bibinfo{journal}{Phys. Rev. Lett.} \textbf{\bibinfo{volume}{105}},
  \bibinfo{pages}{097001} (\bibinfo{year}{2010}).

\bibitem[{\citenamefont{Sasaki et~al.}(2011)\citenamefont{Sasaki, Kriener,
  Segawa, Yada, Tanaka, Sato, and Ando}}]{sasaki2011}
\bibinfo{author}{\bibfnamefont{S.}~\bibnamefont{Sasaki}},
  \bibinfo{author}{\bibfnamefont{M.}~\bibnamefont{Kriener}},
  \bibinfo{author}{\bibfnamefont{K.}~\bibnamefont{Segawa}},
  \bibinfo{author}{\bibfnamefont{K.}~\bibnamefont{Yada}},
  \bibinfo{author}{\bibfnamefont{Y.}~\bibnamefont{Tanaka}},
  \bibinfo{author}{\bibfnamefont{M.}~\bibnamefont{Sato}}, \bibnamefont{and}
  \bibinfo{author}{\bibfnamefont{Y.}~\bibnamefont{Ando}},
  \bibinfo{journal}{Phys. Rev. Lett.} \textbf{\bibinfo{volume}{107}},
  \bibinfo{pages}{217001} (\bibinfo{year}{2011}),
  \urlprefix\url{http://link.aps.org/doi/10.1103/PhysRevLett.107.217001}.

\bibitem[{\citenamefont{Qi et~al.}(2008)\citenamefont{Qi, Hughes, and
  Zhang}}]{qi2008b}
\bibinfo{author}{\bibfnamefont{X.-L.} \bibnamefont{Qi}},
  \bibinfo{author}{\bibfnamefont{T.}~\bibnamefont{Hughes}}, \bibnamefont{and}
  \bibinfo{author}{\bibfnamefont{S.-C.} \bibnamefont{Zhang}},
  \bibinfo{journal}{Phys. Rev. B} \textbf{\bibinfo{volume}{78}},
  \bibinfo{pages}{195424} (\bibinfo{year}{2008}).

\bibitem[{\citenamefont{Kitaev}(2009)}]{kitaev2009}
\bibinfo{author}{\bibfnamefont{A.}~\bibnamefont{Kitaev}}, \bibinfo{journal}{AIP
  Conf. Proc.} \textbf{\bibinfo{volume}{1134}}, \bibinfo{pages}{22}
  (\bibinfo{year}{2009}).

\bibitem[{\citenamefont{Fidkowski and Kitaev}(2010)}]{fidkowski2010}
\bibinfo{author}{\bibfnamefont{L.}~\bibnamefont{Fidkowski}} \bibnamefont{and}
  \bibinfo{author}{\bibfnamefont{A.}~\bibnamefont{Kitaev}},
  \bibinfo{journal}{Phys. Rev. B} \textbf{\bibinfo{volume}{81}},
  \bibinfo{pages}{134509} (\bibinfo{year}{2010}).

\bibitem[{\citenamefont{Fidkowski and Kitaev}(2011)}]{fidkowski2011b}
\bibinfo{author}{\bibfnamefont{L.}~\bibnamefont{Fidkowski}} \bibnamefont{and}
  \bibinfo{author}{\bibfnamefont{A.}~\bibnamefont{Kitaev}},
  \bibinfo{journal}{Phys. Rev. B} \textbf{\bibinfo{volume}{83}},
  \bibinfo{pages}{075103} (\bibinfo{year}{2011}).

\bibitem[{\citenamefont{Turner et~al.}(2011)\citenamefont{Turner, Pollmann, and
  Berg}}]{turner2011}
\bibinfo{author}{\bibfnamefont{A.~M.} \bibnamefont{Turner}},
  \bibinfo{author}{\bibfnamefont{F.}~\bibnamefont{Pollmann}}, \bibnamefont{and}
  \bibinfo{author}{\bibfnamefont{E.}~\bibnamefont{Berg}},
  \bibinfo{journal}{Phys. Rev. B} \textbf{\bibinfo{volume}{83}},
  \bibinfo{pages}{075102} (\bibinfo{year}{2011}).

\bibitem[{\citenamefont{Chen et~al.}(2011)\citenamefont{Chen, Gu, and
  Wen}}]{chen2011}
\bibinfo{author}{\bibfnamefont{X.}~\bibnamefont{Chen}},
  \bibinfo{author}{\bibfnamefont{Z.-C.} \bibnamefont{Gu}}, \bibnamefont{and}
  \bibinfo{author}{\bibfnamefont{X.-G.} \bibnamefont{Wen}},
  \bibinfo{journal}{Phys. Rev. B} \textbf{\bibinfo{volume}{83}},
  \bibinfo{pages}{035107} (\bibinfo{year}{2011}),
  \urlprefix\url{http://link.aps.org/doi/10.1103/PhysRevB.83.035107}.

\bibitem[{\citenamefont{Qi}(2012)}]{qi2012}
\bibinfo{author}{\bibfnamefont{X.}~\bibnamefont{Qi}},
  \bibinfo{howpublished}{e-print arXiv:1202.3983} (\bibinfo{year}{2012}).

\bibitem[{\citenamefont{Ryu and Zhang}(2012)}]{ryu2012}
\bibinfo{author}{\bibfnamefont{S.}~\bibnamefont{Ryu}} \bibnamefont{and}
  \bibinfo{author}{\bibfnamefont{S.-C.} \bibnamefont{Zhang}},
  \bibinfo{howpublished}{e-print arXiv:1202.4484} (\bibinfo{year}{2012}).

\bibitem[{\citenamefont{Yao and Ryu}(2012)}]{yao2012}
\bibinfo{author}{\bibfnamefont{H.}~\bibnamefont{Yao}} \bibnamefont{and}
  \bibinfo{author}{\bibfnamefont{S.}~\bibnamefont{Ryu}},
  \emph{\bibinfo{title}{Interaction effect on topological classification of
  superconductors in two dimensions}}, \bibinfo{howpublished}{e-print
  arXiv:1202.5805} (\bibinfo{year}{2012}).

\bibitem[{\citenamefont{Chen et~al.}()\citenamefont{Chen, Gu, Liu, and
  Wen}}]{chen2011b}
\bibinfo{author}{\bibfnamefont{X.}~\bibnamefont{Chen}},
  \bibinfo{author}{\bibfnamefont{Z.-C.} \bibnamefont{Gu}},
  \bibinfo{author}{\bibfnamefont{Z.-X.} \bibnamefont{Liu}}, \bibnamefont{and}
  \bibinfo{author}{\bibfnamefont{X.-G.} \bibnamefont{Wen}},
  \emph{\bibinfo{title}{Symmetry protected topological orders and the
  cohomology class of their symmetry group}}, \bibinfo{howpublished}{e-print
  arXiv:1106.4772 (2011)}.

\bibitem[{\citenamefont{Gu and Wen}()}]{gu2012}
\bibinfo{author}{\bibfnamefont{Z.-C.} \bibnamefont{Gu}} \bibnamefont{and}
  \bibinfo{author}{\bibfnamefont{X.-G.} \bibnamefont{Wen}},
  \emph{\bibinfo{title}{Symmetry-protected topological orders for interacting
  fermions -- fermionic topological non-linear sigma-models and a group
  super-cohomology theory}}, \bibinfo{howpublished}{e-print arXiv:1201.2648
  (2012)}.

\bibitem[{\citenamefont{Zhang et~al.}(1989)\citenamefont{Zhang, Hansson, and
  Kivelson}}]{zhang1989}
\bibinfo{author}{\bibfnamefont{S.~C.} \bibnamefont{Zhang}},
  \bibinfo{author}{\bibfnamefont{T.~H.} \bibnamefont{Hansson}},
  \bibnamefont{and} \bibinfo{author}{\bibfnamefont{S.}~\bibnamefont{Kivelson}},
  \bibinfo{journal}{Phys. Rev. Lett.} \textbf{\bibinfo{volume}{62}},
  \bibinfo{pages}{82} (\bibinfo{year}{1989}).

\bibitem[{\citenamefont{Wen and Niu}(1990)}]{wen1990b}
\bibinfo{author}{\bibfnamefont{X.~G.} \bibnamefont{Wen}} \bibnamefont{and}
  \bibinfo{author}{\bibfnamefont{Q.}~\bibnamefont{Niu}},
  \bibinfo{journal}{Phys. Rev. B} \textbf{\bibinfo{volume}{41}},
  \bibinfo{pages}{9377} (\bibinfo{year}{1990}),
  \urlprefix\url{http://link.aps.org/doi/10.1103/PhysRevB.41.9377}.

\bibitem[{\citenamefont{Zhang}(1992)}]{zhang1992}
\bibinfo{author}{\bibfnamefont{S.~C.} \bibnamefont{Zhang}},
  \bibinfo{journal}{Int. J. Mod. Phys. B} \textbf{\bibinfo{volume}{6}},
  \bibinfo{pages}{25} (\bibinfo{year}{1992}).

\bibitem[{\citenamefont{Qi et~al.}(2009{\natexlab{b}})\citenamefont{Qi, Li,
  Zang, and Zhang}}]{qi2009}
\bibinfo{author}{\bibfnamefont{X.-L.} \bibnamefont{Qi}},
  \bibinfo{author}{\bibfnamefont{R.}~\bibnamefont{Li}},
  \bibinfo{author}{\bibfnamefont{J.}~\bibnamefont{Zang}}, \bibnamefont{and}
  \bibinfo{author}{\bibfnamefont{S.-C.} \bibnamefont{Zhang}},
  \bibinfo{journal}{Science} \textbf{\bibinfo{volume}{323}},
  \bibinfo{pages}{1184} (\bibinfo{year}{2009}{\natexlab{b}}).

\bibitem[{\citenamefont{Peccei and Quinn}(1977)}]{peccei1977}
\bibinfo{author}{\bibfnamefont{R.~D.} \bibnamefont{Peccei}} \bibnamefont{and}
  \bibinfo{author}{\bibfnamefont{H.~R.} \bibnamefont{Quinn}},
  \bibinfo{journal}{Phys. Rev. Lett.} \textbf{\bibinfo{volume}{38}},
  \bibinfo{pages}{1440} (\bibinfo{year}{1977}),
  \urlprefix\url{http://link.aps.org/doi/10.1103/PhysRevLett.38.1440}.

\bibitem[{\citenamefont{Wilczek}(1987)}]{wilczek1987}
\bibinfo{author}{\bibfnamefont{F.}~\bibnamefont{Wilczek}},
  \bibinfo{journal}{Phys. Rev. Lett.} \textbf{\bibinfo{volume}{58}},
  \bibinfo{pages}{1799} (\bibinfo{year}{1987}).

\bibitem[{\citenamefont{Hansson et~al.}(2011)\citenamefont{Hansson, Karlhede,
  and Sato}}]{hansson2011}
\bibinfo{author}{\bibfnamefont{T.}~\bibnamefont{Hansson}},
  \bibinfo{author}{\bibfnamefont{A.}~\bibnamefont{Karlhede}}, \bibnamefont{and}
  \bibinfo{author}{\bibfnamefont{M.}~\bibnamefont{Sato}},
  \bibinfo{howpublished}{e-print arXiv:1105.5031} (\bibinfo{year}{2011}).

\bibitem[{\citenamefont{Wang et~al.}(2011)\citenamefont{Wang, Qi, and
  Zhang}}]{wang2011}
\bibinfo{author}{\bibfnamefont{Z.}~\bibnamefont{Wang}},
  \bibinfo{author}{\bibfnamefont{X.-L.} \bibnamefont{Qi}}, \bibnamefont{and}
  \bibinfo{author}{\bibfnamefont{S.-C.} \bibnamefont{Zhang}},
  \bibinfo{journal}{Phys. Rev. B} \textbf{\bibinfo{volume}{84}},
  \bibinfo{pages}{014527} (\bibinfo{year}{2011}).

\bibitem[{\citenamefont{Ryu et~al.}(2012)\citenamefont{Ryu, Moore, and
  Ludwig}}]{ryu2012b}
\bibinfo{author}{\bibfnamefont{S.}~\bibnamefont{Ryu}},
  \bibinfo{author}{\bibfnamefont{J.~E.} \bibnamefont{Moore}}, \bibnamefont{and}
  \bibinfo{author}{\bibfnamefont{A.~W.~W.} \bibnamefont{Ludwig}},
  \bibinfo{journal}{Phys. Rev. B} \textbf{\bibinfo{volume}{85}},
  \bibinfo{pages}{045104} (\bibinfo{year}{2012}),
  \urlprefix\url{http://link.aps.org/doi/10.1103/PhysRevB.85.045104}.

\bibitem[{\citenamefont{Witten}(1982)}]{witten1982}
\bibinfo{author}{\bibfnamefont{E.}~\bibnamefont{Witten}},
  \bibinfo{journal}{Physics Letters B} \textbf{\bibinfo{volume}{117}},
  \bibinfo{pages}{324 } (\bibinfo{year}{1982}), ISSN \bibinfo{issn}{0370-2693}.

\bibitem[{\citenamefont{Qi et~al.}(2010)\citenamefont{Qi, Hughes, and
  Zhang}}]{qi2010b}
\bibinfo{author}{\bibfnamefont{X.-L.} \bibnamefont{Qi}},
  \bibinfo{author}{\bibfnamefont{T.~L.} \bibnamefont{Hughes}},
  \bibnamefont{and} \bibinfo{author}{\bibfnamefont{S.-C.} \bibnamefont{Zhang}},
  \bibinfo{journal}{Phys. Rev. B} \textbf{\bibinfo{volume}{81}},
  \bibinfo{pages}{134508} (\bibinfo{year}{2010}).

\bibitem[{\citenamefont{Vollhardt and W\"{o}lfle}(1990)}]{vollhardt1990}
\bibinfo{author}{\bibfnamefont{D.}~\bibnamefont{Vollhardt}} \bibnamefont{and}
  \bibinfo{author}{\bibfnamefont{P.}~\bibnamefont{W\"{o}lfle}},
  \emph{\bibinfo{title}{The Superfluid Phases of Helium 3}}
  (\bibinfo{publisher}{Taylor and Francis, USA}, \bibinfo{year}{1990}).

\bibitem[{\citenamefont{Zhang and Hu}(2001)}]{zhang2001}
\bibinfo{author}{\bibfnamefont{S.~C.} \bibnamefont{Zhang}} \bibnamefont{and}
  \bibinfo{author}{\bibfnamefont{J.~P.} \bibnamefont{Hu}},
  \bibinfo{journal}{Science} \textbf{\bibinfo{volume}{294}},
  \bibinfo{pages}{823} (\bibinfo{year}{2001}).

\bibitem[{\citenamefont{Callan and Harvey}(1985)}]{callan1985}
\bibinfo{author}{\bibfnamefont{C.~G.} \bibnamefont{Callan}} \bibnamefont{and}
  \bibinfo{author}{\bibfnamefont{J.~A.} \bibnamefont{Harvey}},
  \bibinfo{journal}{Nucl. Phys. B} \textbf{\bibinfo{volume}{250}},
  \bibinfo{pages}{427} (\bibinfo{year}{1985}).

\bibitem[{\citenamefont{Teo and Kane}(2010)}]{teo2010}
\bibinfo{author}{\bibfnamefont{J.~C.~Y.} \bibnamefont{Teo}} \bibnamefont{and}
  \bibinfo{author}{\bibfnamefont{C.~L.} \bibnamefont{Kane}},
  \bibinfo{journal}{Phys. Rev. B} \textbf{\bibinfo{volume}{82}},
  \bibinfo{pages}{115120} (\bibinfo{year}{2010}),
  \urlprefix\url{http://link.aps.org/doi/10.1103/PhysRevB.82.115120}.

\bibitem[{\citenamefont{Alvarez-Gaume and Witten}(1983)}]{alvarez-gaume1985}
\bibinfo{author}{\bibfnamefont{L.}~\bibnamefont{Alvarez-Gaume}}
  \bibnamefont{and} \bibinfo{author}{\bibfnamefont{E.}~\bibnamefont{Witten}},
  \bibinfo{journal}{Nucl. Phys. B} \textbf{\bibinfo{volume}{234}},
  \bibinfo{pages}{269} (\bibinfo{year}{1983}).

\end{thebibliography}
\end{document}